\documentstyle[preprint,epsfig,aps]{revtex}

\begin{document}
\title{Quasiparticle Random Phase Approximation with an optimal
       Ground State}

\author{F. \v Simkovic$^{a)}$, M. \v Smotl\'ak$^{a,b)}$,
        and A. A. Raduta$^{c,d)}$}
\address{$^{a)}$Dept. of Nuclear Physics, Comenius University,
Mlynsk\'a dol., pav. F1, SK-842 15 Bratislava, Slovakia\\
$^{b)}$ Joint Institute for Nuclear Research, 141980 Dubna, Russia\\
$^{c)}$Institute of Physics and Nuclear Engineering, 
Bucharest, POB MG6, Romania\\
$^{d)}$Dept. of Theoretical Physics and
Mathematics, Bucharest University, POB MG11, Romania}
\date{\today}
\maketitle

\begin{abstract}
A new Quasiparticle Random Phase Approximation approach is presented. 
The corresponding ground state is variationally determined
and exhibits a minimum energy. New 
solutions for the ground state, some  with spontaneously 
broken symmetry, of a solvable Hamiltonian are found. 
A non-iterative procedure to solve the non-linear QRPA equations is
used and thus all possible solutions are found. These are compared with
the exact results as well as with the solutions provided by other approaches.

\noindent 
PACS number(s):21.60.Jz,21.60.Fw,21.60.Ev,23.40.Hc
\end{abstract}

The random phase approximation (RPA) and its 
quasiparticle (qp) generalization (QRPA) are
powerful tools for describing the  collective degrees
of freedom of many-fermion systems in various branches of physics
like nuclear physics, solid state, plasma physics, atomic clusters.

Usually the  RPA and QRPA are presented by making use of
boson expansion techniques. If the one body transition operator is
expressed linearly in bosons,  the many body Hamiltonian with the
two body interaction included becomes a quadratic polynomial in bosons
describing harmonic motion of the selected degree of freedom.
For the evident reasons the resulting approach is called a 
quasiboson approximation (QBA). 
The main drawback of  this approximation scheme is that the
QRPA (henceforth called as the standard QRPA) 
exhibits a collapse for a critical value of the
strength of the attractive particle-particle force, i.e. the lowest 
eigenvalue becomes imaginary. In the case of the many body Hamiltonians
used for the description of the transition rates of single and double
beta decays the realistic proton-neutron force is expected to be
close to its critical value. Due to this feature many
theoretical works have been devoted to improving the standard 
QRPA by removing the ground state instabilities
\cite{srvf00,radu00,rad20,hirm96,hirm60,qeq3,qeq1,samb97,dang99,samb99,mahi00}.

It is a common practice to investigate the validity of different
approximation schemes within exactly solvable models, which
describe the gross properties of the fermion many-body 
systems 
\cite{srvf00,radu00,rad20,hirm96,hirm60,qeq1,samb97,hag00,engel}.
Recently, it has been shown that the QRPA does not develop any collapse
and is in good agreement with exact results, if  
the Pauli exclusion  principle (PEP) is  consistently implemented 
in this approach \cite{srvf00}.  However, there is another group of studies 
saying that the collapse of the QRPA indicates a
phase transition, i.e., a rearrangement of the
nuclear ground state \cite{radu00,rad20,hirm60,engel}. 
In Ref.\cite{radu00}, in order to avoid the QRPA collapse, a 
new static ground state is defined by means of a semiclassical approach. 
A new collective mode, 
for a many-body system with proton-neutron pairing interaction, 
has been found beyond the point where the QRPA breaks down. This new solution
requires the restoration of the PEP, objective that can be touched
through a boson expansion technique. We note that
the main difference between the  semiclassical 
and the QRPA solutions consists in the description of the 
ground state. 

In this letter we  point out  new features for a 
many fermion system by improving the
QRPA description of the ground state. A new ansatz for the QRPA wave functions 
is proposed, which allows us to
minimize the QRPA ground state energy and, at the same time, to diagonalize
the QRPA equations. The method proposed will be conventionally called as
QRPA with an optimal ground state. One hopes that the present approach
is a suitable tool for a realistic treatment of many body systems.   
It is worth mentioning that the correct description of the ground state 
of a many-body system is a steady interesting subject which 
has attracted
the efforts of many groups \cite{rad20,hirm60,samb99,samb60}.

For the sake of simplicity we consider 
the proton-neutron monopole Lipkin Hamiltonian  which is exactly solvable 
\cite{srvf00,radu00,hirm96,samb97,samb99}
\begin{equation}
H_F = \epsilon ({\hat N}_p + {\hat N}_n) + 
\lambda_1 A^\dagger A + \lambda_2 ( A^\dagger A^\dagger + A A),
\label{eq:1}
\end{equation}                                                    
where ${\hat N}_p$ (${\hat N}_n$) and $A^\dagger$
 are the proton (neutron) number and 
proton-neutron pair qp operators,
respectively \cite{radu00}. The spaces of single particle states
associated to the proton and neutron systems, respectively, are
restricted to a single j-shell.
The model Hamiltonian can 
be obtained from a Hamiltonian which in the particle representation, 
consists of pairing interaction for alike nucleons, and a
monopole-monopole 
proton-neutron two body interaction 
of particle--hole and particle--particle types
\cite{hirm96,samb97,samb99,srvf00,radu00}. 
For the single j-shell, considered here, one obtains:
\begin{eqnarray}
 \lambda_1 &=&  4\Omega [ \chi ( u^2_p v^2_n + v^2_p u^2_n ) - 
                 \kappa (u^2_p u^2_n + v^2_p v^2_n) ],
\nonumber \\
\lambda_2 &=& 4\Omega (\chi + \kappa ) u_p v_p u_n v_n 
~~~\Omega=j+\frac{1}{2},
\label{eq:2}
\end{eqnarray}
where $u_p, v_p$ and $u_n, v_n$  are the coefficients of the
Bogoliubov-Valatin transformation for protons and neutrons, respectively.
The parameters $\chi$ and $\kappa$ are the
strengths of particle--hole and particle--particle 
proton-neutron interactions, respectively.
We consider the boson mapping of the model Hamiltonian, following the
Marumori recipe, i.e. the boson representation $H_B$ is chosen so that
its matrix elements (m.e.) between boson states are equal  to the m.e.
of $H_F$ between the corresponding many fermion states \cite{ring}. The final
result is \cite{samb97,samb99}:
\begin{eqnarray}
H_B&=&\alpha_{11}B^+B + 
   \alpha_{02}(B^+B^+ + BB) + \alpha_{22}B^+B^+BB 
\nonumber\\
&+& 
   \alpha_{13}(B^+BBB + B^+B^+B^+B),
\label{eq:3}
\end{eqnarray}
with 
\begin{eqnarray}
\alpha_{11} &=& (2\varepsilon+\lambda_1),\; 
\alpha_{02}=\lambda_2\sqrt{(1-\frac{1}{2\Omega})}, \;
\alpha_{22} = -\frac{\lambda_1}{2\Omega}, 
\nonumber\\
\alpha_{13}&=&\lambda_2\Big[\sqrt{(1-\frac{1}{2\Omega})(1-\frac{1}{\Omega})}-
\sqrt{(1-\frac{1}{2\Omega})}\Big].
\label{eq:4}
\end{eqnarray}
The operator $B^\dagger$ is a creation boson operator, while $B$ is its
hermitian conjugate operator. 

The QRPA treatment of $H_F$ is determined by a RPA-like procedure
applied to $H_B$. This consists in determining the phonon operator 
\begin{equation}
  Q^\dagger = X (B^\dagger + t) - Y (B + t^*), ~~ Q = (Q^\dagger )^\dagger
\label{eq:5}
\end{equation}
as well as the vacuum state $|rpa\rangle$ and the one phonon state $|Q\rangle$
\begin{equation}
 | Q > = Q^\dagger |rpa>, ~~~~~~ Q |rpa> = 0. 
\label{eq:6}
\end{equation}
From the set of solutions depending on the parameters $(t,t^*)$ one
depicts that one for which the state $|rpa\rangle$ has a minimum energy. 
In contradistinction to the usual QRPA, here the phonon operator involves
a C number which results in having a nonvanishing expected value for $Q$,
in the state $|rpa\rangle$, i.e. a "deformed ground state".

The ansatz for the QRPA ground state $|rpa>$, defined as 
vacuum state for $Q$ [see Eq. (\ref{eq:6})], is taken
as follows: 
\begin{equation}
|rpa> = e^{t^*B - t B^\dagger} ~e^{z B^\dagger B^\dagger -z^* BB} |>.
\label{eq:7}
\end{equation}
Here $|>$ denotes the vacuum state for the boson operator B, which might
be the mapping of the uncorrelated BCS ground state.
$t$, $z$ are complex parameters: $t=r e^{i\theta}$, $z=\rho e^{i\phi}$.
The ansatz (\ref{eq:7}) obeys the second eq. (\ref{eq:6}) provided the
following equation holds:
\begin{equation}  
{\cal F}(r,\theta,\rho,\phi) \equiv 
 \sinh{(2\rho)} - \frac{Y}{X} e^{-i\phi} \cosh{(2\rho )} = 0,
\label{eq:8}
\end{equation}
where the function ${\cal F}$ depends implicitly on $r$ and 
$\theta$ variables, by means of $X$ and $Y$ amplitudes.
It is worth to note that the functions describing the limiting cases 
$(\rho,\phi)=(0,0)$ and $(r,\theta)=(0,0)$ were earlier considered
\cite{radu00}
as trial states in a time dependent formalism: 
\begin{equation}
|rpa\rangle=e^{-\frac{r^2}{2}}e^{-tB^{\dagger}},\;
|rpa> = \frac{1}{\sqrt{\cosh{(2\rho )}}} e^{d B^\dagger B^\dagger} |>
\label{eq:9}
\end{equation}
with $d = e^{i\phi} \tanh{(2\rho )}/2$.
Due to this
feature one expects that the present trial function accounts for new
correlations in the ground state. 
The phonon amplitudes $X,Y$ satisfy the QRPA equations 
\begin{eqnarray}
 {\cal A} X + {\cal B} Y &=&  E_{rpa} {\cal U} X, \nonumber \\
 {\cal B}^* X + {\cal A} Y &=& -E_{rpa} {\cal U} Y,
\label{eq:10}
\end{eqnarray}
where the RPA energy $E_{rpa}$ is the excitation energy for the RPA state
$|Q\rangle$, i.e. $E_{rpa}=E_{1} - E_{g.s.}$. 
The RPA matrices ${\cal A},{\cal B},{\cal U}$, are determined in the
usual manner and have the expressions:
\begin{eqnarray}
{\cal A} & \equiv & <rpa| [B, H_B, B^\dagger ] |rpa> \nonumber \\ 
&=& a_{11} + 4 a_{22} v^2 + 6 a_{13} u v \cos{\phi} 
 + r^2 [ 4 a_{22} + 6 a_{13} \cos{(2\theta )} ],  \\
{\cal B}& \equiv & <rpa| [B, H_B, B ] |rpa> \nonumber \\ 
&=& 2 a_{02} + 2 a_{22} u v [ \cos{\phi} + i \sin{\phi} ] 
+ 6 a_{13} v^2 
+ r^2 ( 2 a_{22} [ \cos{(2\theta )} + i \sin{(2\theta )} ] + 6 a_{13} ),
\nonumber \\
{\cal U} & \equiv & <rpa| [B, B^\dagger ] |rpa> = 1. 
\label{eq:11}
\end{eqnarray}
The double commutators are defined as 2$[A,B,C]=[A,[B,C]] +[[A,B],C]$.
Also, the notations $u = \cosh{(2\rho )}$ and $v = \sinh{(2\rho )}$
have been newly introduced.
The matrix ${\cal A}$ is real, while ${\cal B}$
might be a complex number. In this letter we consider only situations where
the solutions for the $X,Y$ amplitudes are real.
The arguments which support our choice is that the resulting phonon state
should be the image, through the boson mapping, of a state in the fermionic
space. However the amplitudes of the
eigenstates of $H_F$, in the basis 
$\{(A^{\dagger})^n\}$,
are real numbers. Thus the possible solutions for $\theta$ and $\phi$
are
$\theta = {n \pi}/{2}$ and $\phi = n \pi$ (n = 1,2,...).
The additional parameters $r$ and $\theta$ entering the QRPA 
eigenvalue problem will be fixed by requiring
that the expectation value of $H_B$ in the
RPA ground state reaches its minimum value $E_{g.s.}$, in the space of
these parameters. One finds 
\begin{eqnarray}
H(r) &\equiv& <rpa| H_B |rpa> = C_0 + C_2 r^2 + C_4 r^4,
\label{eq:12}
\\
C_0 &=& a_{11} v^2 + 2 a_{02} u v \cos{\phi} + 
 a_{22} [u^2 v^2 + 2 v^4 ] 
\nonumber\\
&&+ 6 a_{13} u v^3 \cos{\phi}, \nonumber \\
C_2 &=& a_{11} + 2 a_{02} \cos{(2\theta )}
 + a_{22} [ 4 v^2 + 2 u v \cos{(2 \theta + \phi )} ] 
\nonumber\\
 &&+ 6 a_{13} [ u v \cos{\phi} + v^2 \cos{(2\theta )} ], \nonumber \\
C_4 &=& a_{22} + 2 a_{13} \cos{(2\theta )}.
\label{eq:13}
\end{eqnarray}
Several minima for $E_{g.s.}$ are to be mentioned:\\ 
a) $r = 0$ and $\theta = anything$. This solution corresponds 
to the  RPA ground state given by the second Eq. (\ref{eq:9}).\\
b) $r =\sqrt{- C_2/{2 C_4}}$ ,  $\theta = 0$,$\;$
c) $r =\sqrt{- C_2/{2 C_4}}$, $\theta = \pi/2$,\\
d)  $r =\sqrt{- C_2/{2 C_4}}$, $\theta \ne n \pi/2$ (n=1,2,3 ...).\\
The case d) yields a complex ${\cal B}$ and therefore is not
considered here. The solutions b), c) and d) are specific for
the ansatz (\ref{eq:7}).
If
only the quadratic terms in bosons are considered in  $H_B$ (i.e., $a_{22}, a_{13}=0 $)
then only the solution a) survives. 
The equations (10), (8) and one of a),b),c) are to be simultaneously solved.
This non-linear problem is usually treated  iteratively. However, this
procedure does not guarantees that all solutions of this
system of equations, are found. 
The procedure used in the present paper is as follows. 
First, we insert the solution for $r$ ($r=0$ or $r^2 = - C_2/{2 C_4}$)
in the expressions defining the matrices
${\cal A}$ and ${\cal B}$ by Eqs.(11), (\ref{eq:11}).
Determine the ratio $Y/X$ from any of equations (\ref{eq:10}) 
and insert the result in Eq. (\ref{eq:8}). Then 
 Eq. (\ref{eq:8}) provides $\rho$ which is to be inserted in the
equations for the RPA matrices. With the matrices ${\cal A}, {\cal B}$
fully known one determines
the QRPA eigenvalue and eigenvectors, by solving the equations (10). This
procedure is repeated for each minimum of $E_{g.s.}$
[the cases a), b) and c) mentioned above]. We note that 
solutions with ($\rho$, $\phi$) and ($-\rho$, $\phi\pm\pi$) are
degenerate.

The numerical application is performed for a system of 4 protons and 6
neutrons moving in a $j=\frac{9}{2}$ shell. The strength parameters
defining the model Hamiltonian are $\epsilon=1 MeV$ while the two body interaction
strengths  $\kappa$ and $\chi$  are re-scaled as 
in Refs. \cite{hirm96,samb97}:
$\kappa \rightarrow {\kappa}' \equiv 2\Omega~ \kappa$,
$\chi \rightarrow {\chi}' \equiv 2\Omega ~ \chi$.
We adopt ${\chi}'=0.5$ while $\kappa'$ is considered as a free
parameter in the interval $(0,3) MeV $.
In Table \ref{table.1} we present the results for the QRPA and
ground state energies corresponding to three values of $\kappa'$.
If $\kappa'=0.5$ there are only two solutions of type a),
associated with the ansatz (\ref{eq:9}) (second equation). 
The solution a2) is similar to
that obtained in \cite{srvf00} within the EPP QRPA formalism. The 
solution a1), with small QRPA energy and a very large
value of $E_{g.s.}$, has not been seen before. Its finding is a merit of
the adopted numerical procedure.
For $\kappa'= 1.5, 2.5$ one finds solutions of type
c) which are specific for the new ansatz of the RPA
wave functions. 

\begin{figure}[h]
\centerline{\epsfig{file=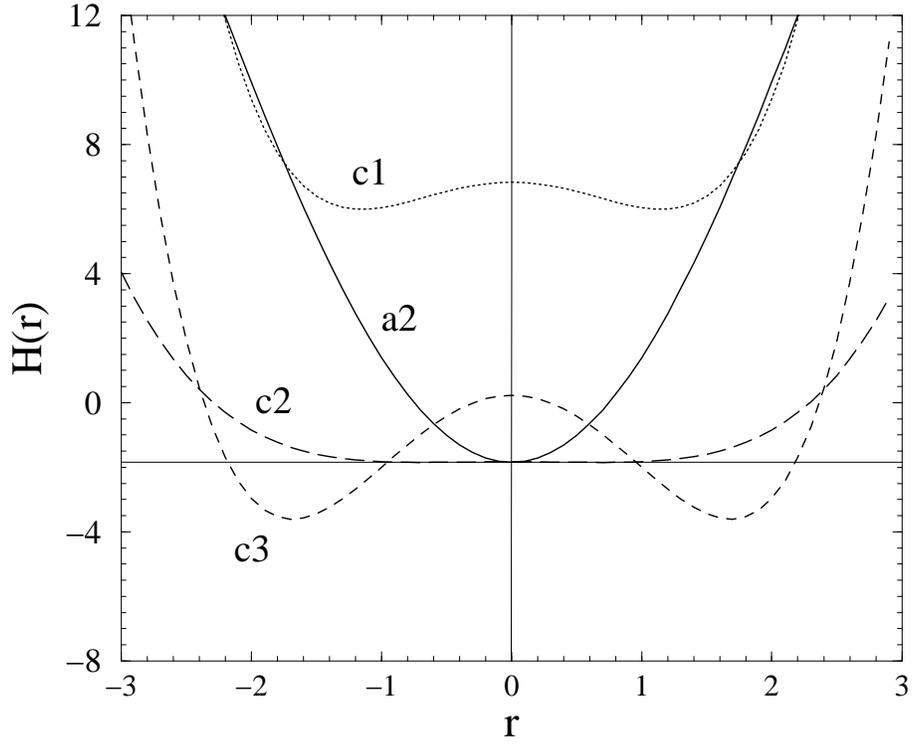,width=12.cm}}
\vspace*{0.5cm}
\caption{ Expectation value of the Hamiltonian $H_B$ as a function
of the parameter $r$. The solutions labeled by a) and c)
are those presented in the text below Eq. (14). The adopted values
of $\chi'$ and $\kappa'$ are $0.5$ and $2.5$, respectively.
}
\label{fig.1}
\end{figure}

In Fig. \ref{fig.1} 
we present the potential energy surface as function of
$r$  for the solutions a2), c1), c2) and c3),
assuming $\kappa'=2.5$. Note that for the solution c)
$H(r)$ has two minima, at $r = \pm \sqrt{-C_2/(2 C_4)}$.

The meaning of the negative r consists of that the result for energy
is not affected if we change the ansatz
by replacing $t$ with $-t$.

\begin{figure}[h]
\centerline{\epsfig{file=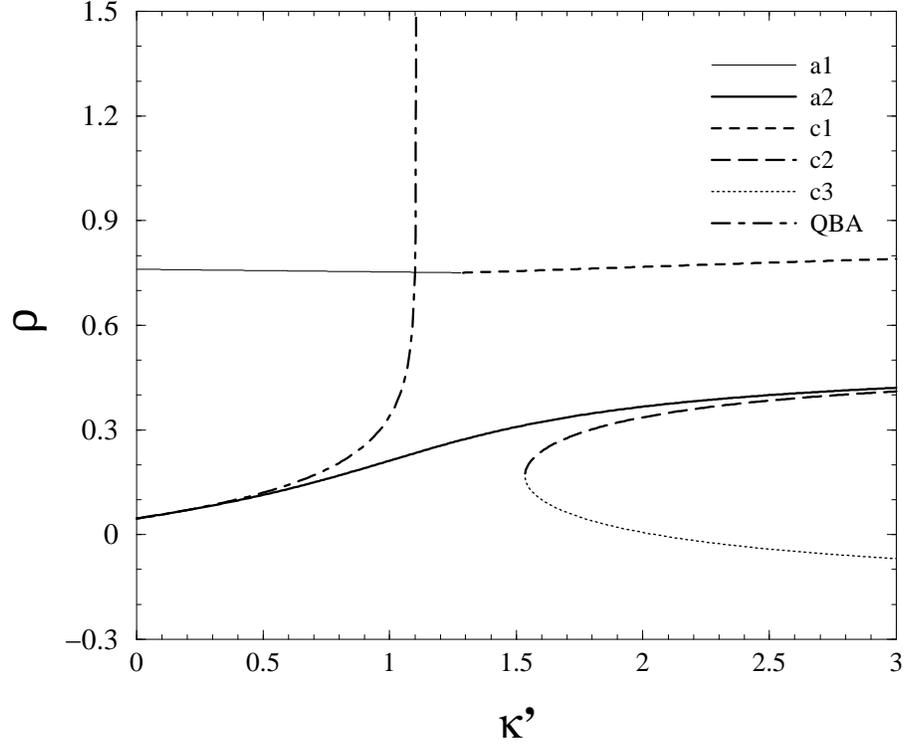,width=12.cm}}
\vspace{0.5cm}
\caption{The parameter $\rho$ of the RPA wave function 
[see Eqs. (7) and (9)] associated
with the solutions  a) and c)  is plotted versus $\kappa'$.
}
\label{fig.2}
\end{figure}

In Fig. \ref{fig.2} the values of $\rho$ corresponding to different 
solutions are presented. Within the QBA,
i.e. the quadratic bosonic Hamiltonian is considered, the value
of $\rho$ is increasing rapidly in the vicinity of the 
QRPA collapse. The  solution a2)
does exist within the whole interval of $\kappa'$.
This is caused by that the PEP principle included 
in the higher order bosonic expansion
Hamiltonian prevents the collapse of the QRPA
\cite{srvf00}. The solution a1) was found
within the range $0.0 \le \kappa' \le 1.29$ and continues further
as c1) solution.
The  solutions c2) and c3) start with $\kappa' = 1.53$ where they have
equal values. For large value of $\kappa'$
the solution c2) is close to a2). It is worthwhile to notice
that in the case of c3), with the lowest value of $E_{g.s}$ and large
$\kappa'$ (see Table \ref{table.1}), the corresponding values of
$\rho$ are rather small. One may say that the corellations in the ground state
are mainly induced by the "deformation" of the system, i.e. the
stationary value of r. For $\kappa' > 2$, $\rho$ is slightly
negative. This branch is obtained by shifting the phase
$\phi$ with $\pi$ which is equivalent to changing the sign of $\rho$ 
but keeping $\phi$ unchanged.

\begin{figure}[h]
\centerline{\epsfig{file=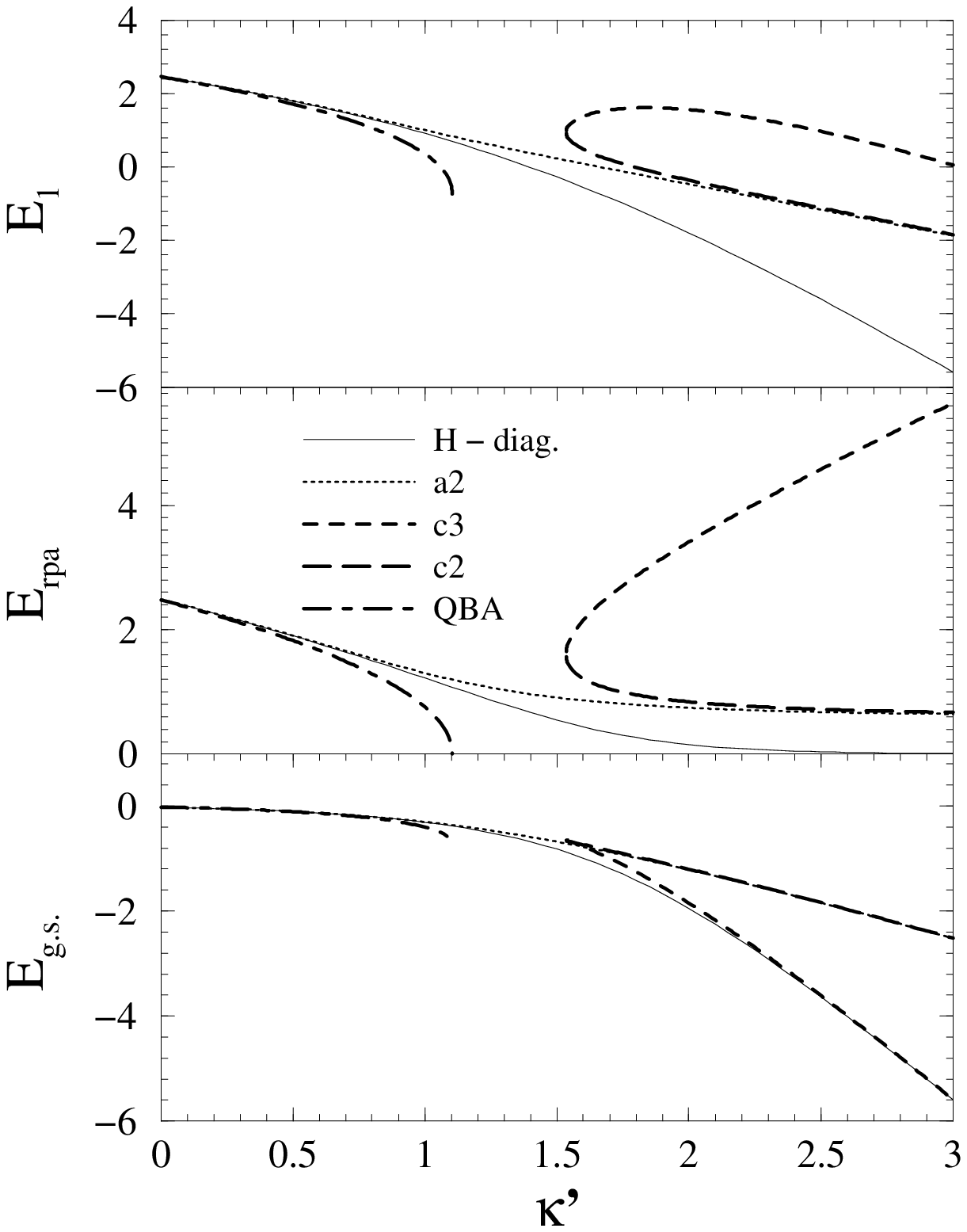,width=14cm}}
\caption{  Energies of the ground state $E_{g.s}$, 
of the first  excited state $E_{1}$ and
the excitation energies $E_{rpa}(= E_1 - E_{g.s}$)
provided by diagonalizing $H_F$, by the standard QRPA and by the
QRPA with an optimal ground state [ a1), c2) and c3)], respectively, are
plotted as function of $\kappa'$. 
}
\label{fig.3}
\end{figure}

In Fig. \ref{fig.3} we plotted  the ground
state energy $E_{g.s}$, the QRPA energy $E_{rpa}$ and
the first excitation energy $E_1$ obtained
by the standard QRPA, QRPA with an optimal ground state
and by diagonalizing $H_F$, as function  of $\kappa'$. Only solutions 
with low ground state energy are drawn. 
One remarks that by including
the higher order boson terms in the Hamiltonian $H_B$
and considering the standard ansatz for $|rpa\rangle$ (see the second Eq. (9))
[solution a2)], there is  no collapse of the QRPA solution.
This solution coincides well with that provided by the
EPP QRPA  formalism \cite{srvf00}. However, for large values of 
$\kappa'$ ($\kappa'\ge 1.6$) this solution approximates poorly the 
exact result for $E_{g.s}$ . 
On other hand the new solution c3) is in an excellent agreement 
with the exact $E_{g.s}$ for this range of $\kappa'$.
In fact, the exact result for $E_{g.s}$ is very well described by matching the
approaches a2) and c3). Despite the fact the c3) approach simulates
excellently the exact result for the ground state energy, the two treatments
predict QRPA energies which deviate from each other by a large amount.

In summary, a new ansatz for the QRPA wave functions was 
proposed. The additional new factor in this ansatz depends on a
complex parameter t which is fixed by requiring that the corresponding
expectation value of the model Hamiltonian $H_B$ is minimum.
The present paper constitute the first attempt in the literature, to determine the QRPA
ground state (the vacuum state for the Q operator) as a solution of a
variational equation.
The numerical application shows clearly new effects caused by
the presence of higher order boson terms in $H_B$ and by the complex
structure of the ansatz for the state $|rpa\rangle$.
Both features mentioned above prevent the QRPA to collapse.
Several solutions are found in the interval of the interaction strength
$\kappa'$ beyond the critical value.
Each solution corresponds to a certain type of minimum in the $(r,\theta)$
variable and by that to a certain symmetry of the wave function.
One may say that the present approach accounts, in an unified fashion, 
for the complementary features described by the EPP QRPA and the semiclassical
formalism, which defines a new ground state for those $\kappa'$
where the standard QRPA fails.


\begin{table}[t]
\caption{The predicted QRPA ($E_{rpa}$) and 
ground state ($E_{g.s}$) energies for 
$\kappa' = 0.5$, $1.5$ and $2.5$, $\chi' = 0.5$
and $j= 9/2$. The parameters determining the RPA  wave functions and  
the ground state energy  are listed as well. 
} 
\label{table.1}
\begin{tabular}{lccccccccc}
Type  &  $\rho$ & $\phi$ & $r^2$ & $2\theta$ & 
$C_0$ & $C_2$ &  $C_4$ &  $E_{g.s.}$ & $E_{rpa}$ \\ 
  &  & & &  &  &  &  &  [MeV] & [MeV] \\ \hline
\multicolumn{10}{c}{  $\kappa'=0.5$, $E^{exact}_{g.s.} = -0.104~MeV$   } \\
 a1  &   0.757 & 0  & 0 & - &  
7.001 &  0.002 &  -0.100 &  7.001 &   0.046 \\
 a2  &    0.114 & $\pi$ & 0 & -  &  
 -0.104 &  3.004  &  -0.100    &  -0.104  &   1.906    \\
\multicolumn{10}{c}{  $\kappa'=1.5$, $E^{exact}_{g.s.} = -0.823~MeV$  } \\
 a2 &   0.308  & $\pi$ & 0 & -  &  
  -0.679  &   3.120  & -0.100  &  -0.679 & 0.908    \\
 c1 &   0.755  & 0 &  0.438  & $\pi$  &  
  7.004  &  -0.249   &  0.284  & 6.950  &  0.024   \\
\multicolumn{10}{c}{  $\kappa'=2.5$,  $E^{exact}_{g.s.} = -3.638~MeV$  } \\
 a2 &   0.400   & $\pi$ & 0 & -  &  
  -1.841  &   3.344   &  -0.100  & -1.841  &   0.675   \\
c1  &    0.780   & 0  &   1.326  & $\pi$  &  
    6.834    &  -1.264    &   0.476  &   5.996  &    0.112     \\
c2   &   0.384    & $\pi$ &  0.081  & $\pi$  &  
  -1.832     &  -0.077    &  0.476   &   -1.835  &   0.719   \\
c3   &   0.041   & 0 &  2.838 & $\pi$  &  
   0.230   &  -2.705    &   0.476  &   -3.609  &    4.580 \\
\end{tabular}
\end{table}


\begin{references}
\bibitem{srvf00}  \v Simkovic F,  Raduta A A,  Veselsk\'y M and  Faessler A
 2000  {\em Phys. Rev. C}  {\bf 61}  044319
\bibitem{radu00} Raduta A A,  Haug O,  \v Simkovic F and  Faessler A
 2000  {\em J. Phys. G}  {\bf 26} 1327; 2000 {\em Nucl. Phys. A} {\bf 671} 255 
\bibitem{rad20} Raduta A A,  Pacearescu L, Baran V,  Sariguren P and 
Moya de Guerra E 2000 {\em Nucl. Phys.  A} {\bf 675}  503 
\bibitem{hirm96} Hirsch J G, Hess P O and  Civitarese O 
 1996 {\em Phys. Rev. C} {\bf 54} 1976; 1997  {\em Phys. Rev. C} {\bf 56}  199 
\bibitem{hirm60}  Hirsch J G,  Hess P O  and  Civitarese O 1999 
   {\em Phys.  Rev. C} {\bf 60} 064303 
\bibitem{qeq3} Hirsch J G,  Civitarese O and Reboiro M, 
 1999  {\em Phys. Rev. C} {\bf 60}   024309 
\bibitem{qeq1} Krmpoti\'c F,  de Passos E J V, Delion D S,  Dukelsky J, 
   and  Schuck P 1998 {\em Nucl. Phys. A} {\bf 637} 295 
\bibitem{samb97}  Sambataro M and  Suhonen J 1997 {\em Phys. Rev. C} {\bf 56}
   782
\bibitem{dang99} Sambataro M  and  Dang N D 1999 {\em Phys. Rev. C} {\bf 59}  1422 
\bibitem{samb99} Sambataro M 1999 {\em  Phys. Rev. C} {\bf 59}  2056 
\bibitem{mahi00} Mariano A and Hirsch J G 1998 {\em Phys. Rev. C} {\bf 57} 3015; 
   1998 {\em  Phys. Rev. C} {\bf 58}  2736; 2000  {\em  Phys. Rev. C} {\bf 61}  054301 
\bibitem{pass98} de Passos E J V,  de Toledo Piza A F R and   Krmpoti\'c F 1998 
   {\em Phys. Rev. C} {\bf 58}  1841 
\bibitem{samb60} Sambataro M 1999 {\em Phys. Rev. C} {\bf 60} 064320 
\bibitem{hag00} Hagino K and Bertsch G 2000 {\em Phys. Rev. C} {\bf 61}    024307 
\bibitem{engel} Engel J,  Pittel S, Stoitsov M, Vogel P and  Dukelsky J,
   {\em  Phys. Rev. C} {\bf 55} 1781 
\bibitem{ring} P. Ring and P. Schuck, {\it The Nuclear Many-Body System}
 (Springer, New York, 1980). 
\end{references}
\end{document}